# Optimizing the Performance of Reactive Molecular Dynamics Simulations for Multi-Core Architectures

Hasan Metin Aktulga[1], Christopher Knight[2], Paul Coffman[2], Kurt A. O'Hearn[1], Tzu-Ray Shan[3], and Wei Jiang[2]

[1]Michigan State University
[2]Argonne Leadership Computing Facility
[3]Sandia National Laboratories (currently, Materials Design Inc.)

June 23, 2017

**Abstract**

Reactive molecular dynamics simulations are computationally demanding. Reaching spatial and temporal scales where interesting scientific phenomena can be observed requires efficient and scalable implementations on modern hardware. In this paper, we focus on optimizing the performance of the widely used LAMMPS/ReaxC package for multi-core architectures. As hybrid parallelism allows better leverage of the increasing on-node parallelism, we adopt thread parallelism in the construction of bonded and nonbonded lists, and in the computation of complex ReaxFF interactions. To mitigate the I/O overheads due to large volumes of trajectory data produced and to save users the burden of post-processing, we also develop a novel *in-situ* tool for molecular species analysis. We analyze the performance of the resulting ReaxC-OMP package on Mira, an IBM Blue Gene/Q supercomputer. For PETN systems of sizes ranging from 32 thousand to 16.6 million particles, we observe speedups in the range of 1.5-4.5×. We observe sustained performance improvements for up to 262,144 cores (1,048,576 processes) of Mira and a weak scaling efficiency of 91.5% in large simulations containing 16.6 million particles. The *in-situ* molecular species analysis tool incurs only insignificant overheads across various system sizes and run configurations.

## 1 Introduction

Molecular Dynamics (MD) simulations have become an increasingly important computational tool for a range of scientific disciplines including, but not limited to, chemistry, biology, and materials science. To examine the microscopic properties of molecular systems of millions of particles for several nanoseconds (and possibly microseconds), it is crucial to have a computationally inexpensive, yet sufficiently accurate, interatomic potential. Several popular molecular force fields are readily available for modeling liquids, proteins, and materials (e.g. Charmm [1], Amber [2], and OPLS [3]). The computational efficiency of these models can be largely attributed to defining fixed bonding topologies within (and between) molecules, fixed partial charges, and the use of relatively simple functions to model the interatomic potential. While appropriate for many systems and problems, the use of fixed bonding topologies and charges prevents these classical MD models from exploring processes involving chemical reactions or responses from environmental effects, which may be critical to properly understanding a process of interest.

Instead of resorting to computationally expensive quantum mechanical alternatives that explicitly treat the electronic degrees of freedom and therefore are restricted to modeling systems of only a few thousand atoms, one can employ simulation methods that include some degree of variable bond topology (e.g. multi-state methods [4, 5]) or force fields that do not define a fixed bonding topology. This latter class of force fields are called bond order potentials, examples of which include ReaxFF [6, 7], COMB [8, 9] and AIREBO [10] potentials. The goal of all such reactive methodologies and force fields is to model reactive systems at time and length scales that far surpass those currently practical for electronic structure methods, complementing these more accurate quantum mechanics based models. Efficient implementations of such reactive methodologies are crucial to address challenging scientific questions.

In this paper, we focus on ReaxFF, a bond order potential that has been widely used to study chemical reactivity in a wide range of systems. The PuReMD software [11, 12, 13] and the LAMMPS/ReaxC package [14] (which is based on PuReMD) provide efficient, open-source implementations of the ReaxFF model that are currently being used by a large community of researchers. PuReMD and LAMMPS/ReaxC have introduced novel algorithms and data structures to achieve high performance in force computations while retaining a small memory footprint [11, 12]. The ability for a large community of researchers to efficiently carry out such simulations is becoming even more important as algorithms for the efficient fitting of ReaxFF models have been made available recently [15, 16, 17, 18].

Just like computational methods to accurately and efficiently model atomistic systems have evolved over time, so too have the architectures of high performance computing (HPC) systems on which these simulations are executed. Due to the unsustainable levels of power consumption implied by high clock rates, we witnessed the emergence of multi-core architectures over the past decade. Hybrid parallelism (typically in the form of MPI/OpenMP) allows HPC applications to better leverage the increasing on-node parallelism on current generation platforms, such as Intel Xeon, Xeon Phi, and IBM BlueGene/Q. In this paper, we present the techniques and data structures that we used to develop a hybrid parallel ReaxFF software, where the construction of bonded and non-bonded lists and computation of complex interactions have been re-designed to efficiently leverage thread parallelism. Another important trend on HPC systems is the widening gap between their computational power and I/O capabilities. To mitigate the I/O overheads due to large volumes of ReaxFF trajectory data produced and save users the burden of post-processing, we also developed a novel *in-situ* tool for molecular species analysis. We analyze the performance of the resulting ReaxC-OMP package in LAMMPS [19] on Mira, an IBM Blue Gene/Q supercomputer. For system sizes ranging from 32 thousand to 16.6 million particles, we observe speedups in the range of 1.5-4.5× using the new hybrid parallel implementation. Sustained performance improvements have also been observed for up to 1,048,576 processes in larger simulations. We also demonstrate that the *in-situ* molecular species analysis tool incurs only modest overheads depending on the run configuration.

## 2 Background and Motivation

In this section we give a brief overview of ReaxFF's computational workflow, and discuss the specific aspects of ReaxFF that make a hybrid parallel implementation compelling from a performance standpoint.

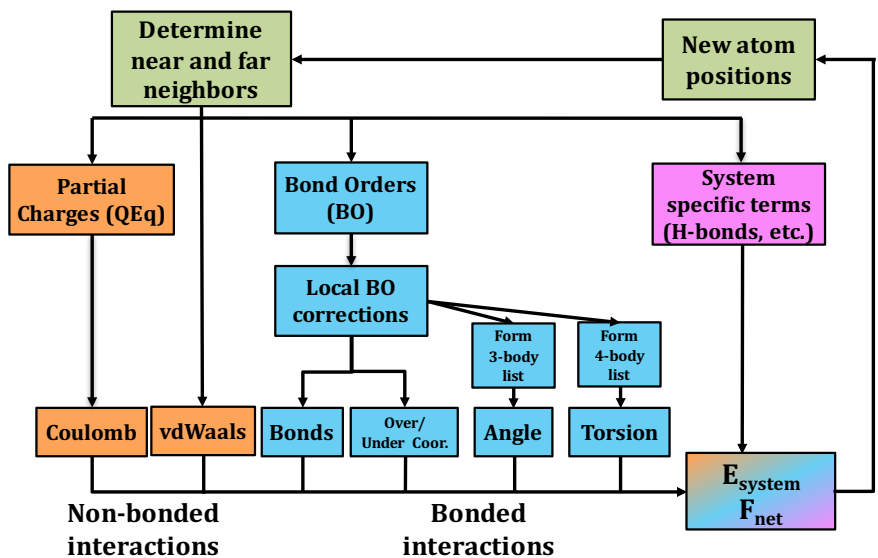

Figure 1: ReaxFF's computational workflow includes bonded, non-bonded and system specific interactions, each with different physical formulations and cut-offs. Figure has been adapted from [7].

## 2.1 ReaxFF Overview and Workflow

ReaxFF [6, 7] replaces the harmonic bonds of molecular models with bond orders, and several partial energy terms that are dependent on inter-atomic distances. Accurately modeling chemical reactions, while avoiding discontinuities on the potential energy surface, however, requires interactions with more complex mathematical formulations than those found in typical molecular models. In a reactive environment where atoms often do not achieve their optimal coordination numbers, ReaxFF requires additional modeling abstractions such as lone pair, over/under-coordination, and 3-body and 4-body conjugation potentials, which introduce significant implementation complexity and computational cost. The satisfaction of valencies, which is explicitly performed in molecular models, necessitates many-body calculations in ReaxFF. An important part of the ReaxFF method is the charge equilibration procedure which tries to approximate the partial charges on atoms using suitable charge models [20, 21]. Charge equilibration is mathematically formulated as the solution of a large sparse linear system of equations, and it needs to be performed accurately at each time-step to ensure proper conservation of energy. As a whole, the ReaxFF approach allows reactive phenomena to be modeled with atomistic resolution in a molecular dynamics framework. Consequently, ReaxFF can overcome many of the limitations inherent to conventional molecular simulation methods, while retaining, to a great extent, the desired scalability.

Figure 1 depicts the various ReaxFF interactions and summarizes the work flow of a simulation. The work flow in ReaxFF is to compute various atomic interaction functions (bonds, lone pair, over-/under-coordination, valance angles, torsions, van der Waals and Coulomb) for the local atomic system (including ghost particles) and then sum various force contributions at the individual atomic level to obtain the net force on each atom for a given time step. Potential energies are computed at the system level.

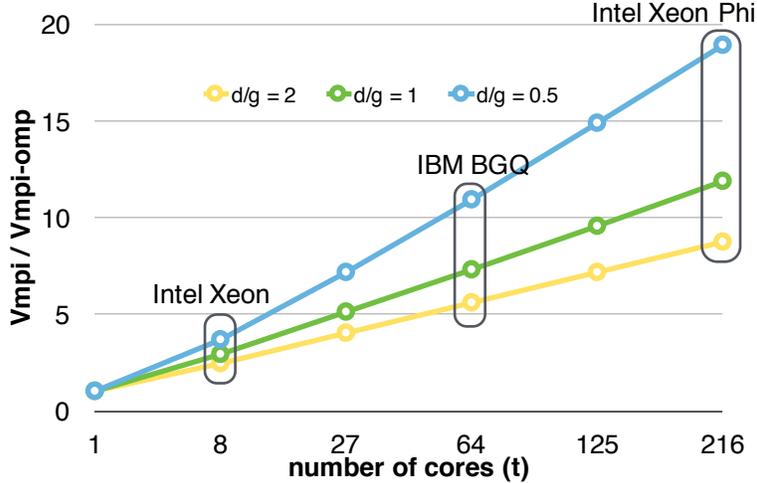

Figure 2: Ratio of the ghost region to the actual simulation domain in MPI-only vs. MPI/OpenMP parallelization under the weak-scaling scenario with increasing number of cores on a single node.

## 2.2 Motivation for a Hybrid Implementation

Parallelization through spatial decomposition, where each MPI process is assigned to a specific region of the simulation box, is the most commonly used approach in MD software, including LAMMPS [19] and PuReMD [12]. With spatial decomposition, the computation of bonded and short-ranged interactions requires the exchange of atom position information near process boundaries, *a.k.a* the ghost region. Communications associated with the ghost regions are an important bottleneck against scalability. Thus, reducing these communication overheads has been the subject of several studies [12, 19, 22, 23, 24]. Note that the amount of ghost region communications and the required data duplication is proportional to the surface area of the domain owned by an MPI process [19]. A hybrid MPI/OpenMP implementation can help in this regard because it naturally reduces i) the number of domain partitions for a given node count, ii) the volume of data exchanges between MPI processes and iii) the redundant computations at the ghost regions (if any).

Below, we quantify this effect with a simple example where we assume a homogeneous (or random) distribution of atoms in a simulation box. The volume ratio of the ghost region to the original simulation domain in an MPI-only vs. MPI/OpenMP implementation would then reflect the relative communication and computation overheads in both schemes. For simplicity, let $d$ denote a dimension of the *cubic* region $V$ assigned to a process (i.e., $d = \sqrt[3]{V}$), $g$ be the thickness of the ghost region (which is typically determined by the largest interaction cutoff distance), $t$ be the number of threads available for parallelization on a node (for simplicity of presentation, we assume $t = c^3$ for some integer $c \geq 1$), and $n$ be the number of nodes used in a computation. Then the total ghost volume in MPI-only and MPI/OpenMP hybrid implementations would respectively be:

$$\begin{aligned} V_{mpi} &= nc^3\left((d+g)^3 - d^3\right) \\ &= n\left((cd+cg)^3 - (cd)^3\right) \\ V_{mpi-omp} &= n\left((cd+g)^3 - (cd)^3\right) \end{aligned} \quad (1)$$

Figure 2 shows the relative volume of the ghost region to the original simulation domain with increasing degree of on-node thread parallelism $t$ and various $\frac{d}{g}$ ratios under a weak-scaling scenario.

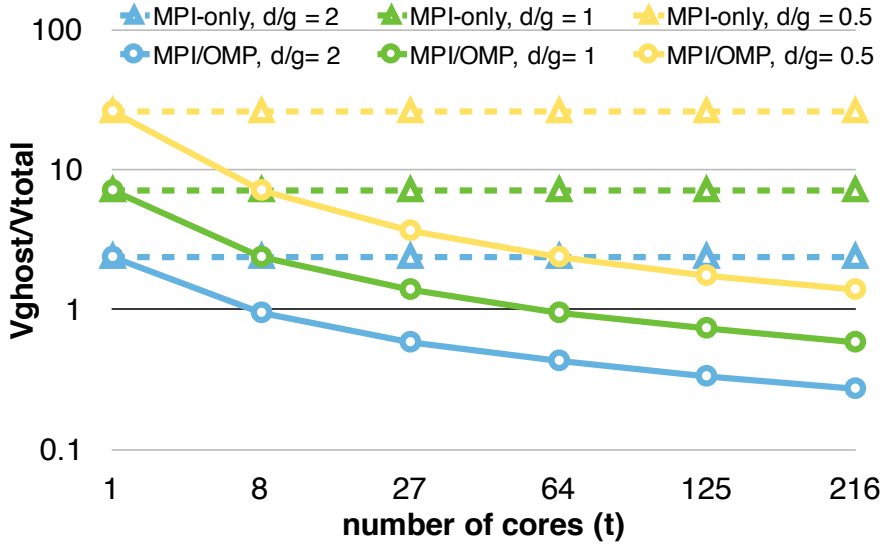

Figure 3: Relative ghost region volumes in MPI-only vs. MPI/OpenMP parallelized molecular dynamics simulations with spatial decomposition. We mark the core counts for typical multi-core and many-core processors available today.

Using MPI-only parallelization, the ratio of the ghost region to the actual simulation domain is constant and is significantly high for low values of $\frac{d}{g}$. Under MPI/OpenMP parallelization, there can be a single partition per node and therefore the relative volume of the ghost region decreases as the degree of on-node thread parallelism increases. As we show in Figure 3, the reduction in the total ghost region volume can be significant for modern architectures like the IBM BG/Q and Intel Xeon Phi systems.

It may be argued that in an MPI-only implementation, expensive inter-process communications may be turned into intra-node communications by mapping the $c^3$ nearby MPI processes onto the same node using, for example, topology aware mapping techniques [25]. As a result, for most classical molecular dynamics models, increased ghost region volume would result in memory overheads, but may not incur significant computational overheads except for building the neighbor lists. In such cases, a hybrid parallel implementation may not yield significant gains for small scale computations, but for capability scale simulations on large supercomputers, leveraging thread parallelism will still be very important.

For ReaxFF computations though, efficiently leveraging hybrid parallelism is crucial in terms of performance for two unique reasons. First, the dynamic nature of bonds in ReaxFF and the presence of valence and dihedral interactions that straddle long distances into process boundaries require a significant number of bonded computations to be repeated in the ghost regions of multiple processes. Unlike most classical MD packages, the computational expense of bonded interactions in ReaxFF are comparable to that of non-bonded interactions [11] due to the presence of several bond-related partial energy terms, each requiring high numbers of arithmetic operations. For instance, using the TATB benchmark in LAMMPS, we observe that the ratio of the computational expenses of bonded and non-bonded interactions is approximately 1.5, 60.77% vs. 39.23% to be exact (this ratio will show variations depending on the cutoffs used and the specific system being simulated). Therefore in the strong scaling limit as $d$ gets comparable to or less than $g$, increased ghost region volumes are likely to cause significant (bonded) computational overheads in ReaxFF simulations.

A second reason is the inter-node communication overheads during the charge equilibration

(QEq) procedure [21]. To determine partial charges on each atom, it is necessary to solve a large linear system of equations at each step of the simulation. For this purpose, iterative linear solvers that require at each iteration a forward-backward halo-exchange of partial charges are used [11]. These communications increasingly become a performance bottleneck as the number of MPI ranks increases. While non-blocking communication primitives can be used to overlap communication and sparse matrix computations during QEq, at scale (i.e., when $d$ is small with respect to $g$) computations with local particles are highly likely to involve interactions with ghost particles belonging to several different neighboring processes. Consequently, overlapping communication with computation is not practical when it is most needed. However, hybrid parallelism can significantly reduce the number of MPI ranks and the number of ghost particles that need to be exchanged during the QEq procedure. As a result, it is expected to reduce the onset of communication-related performance bottlenecks.

These two unique aspects of ReaxFF simulations have been our primary motivation for a hybrid parallel implementation. As we show through extensive tests in the performance evaluation section, we achieve significant performance improvements by porting the LAMMPS/ReaxC package to multi-core architectures.

## 3 Algorithms and Implementation

In this section, we focus on enabling efficient thread parallelism for ReaxFF computations. For interested readers, algorithms, data structures and implementation details underlying our MPI-parallel ReaxFF software (PuReMD and LAMMPS/ReaxC package) are presented in detail by Aktulga *et al.* [11, 12].

### 3.1 Thread Parallelization Strategy

Energy and force computations in ReaxFF, although disparate in their mathematical formulations, need to be aggregated in the same global data structures, with those related to forces being uniquely indexed for each (local and ghost) atom. The force computation functions in ReaxC share a general methodology of computing the energies and forces in an atom centered fashion, defining an interaction list for each atom, calculating the force between a given atom and each of its neighbors, and then aggregating the forces on individual atoms and the potential energy of the system. This methodology is implemented as an outer loop over the data structure containing all atoms in the local system, and an inner loop over the neighbors of a given atom where most of the computation takes place (see Algorithm 1 for an example). Performance counters instrumented within each function around these loops identified them as targets for performance improvements via OpenMP multi-threading. The ensuing tuning effort utilized these counters to precisely measure the OpenMP speedups.

Leveraging Newton's third law which states that for every action, there is an equal and opposite reaction, the computational costs in MD can be reduced by half by computing each interaction between a pair of distinct atoms only once. Contrary to the conventional MD approaches though, Anderson *et al.* has shown that redundantly computing these equal and opposite interactions can be more advantageous on massively parallel processors like GPUs [26], as this approach exposes more parallelism and avoids frequent thread synchronizations. However, interactions in the Reax force field are complex mathematical formulations requiring a high number of arithmetic operations, and a redundant computation approach would require computing the three-body and four-body interactions three and four times, respectively. Therefore we evaluate all interactions (pairwise, three-body and four-body) once and apply the resulting forces to all atoms involved in the interaction.

---
**Algorithm 1** Pairwise force computation
---
**Input:** Atom list and positions: atoms
**Output:** Potential energy, forces (partial): gEnergy, gForce
 1: **for** (int $i = 0$; $i < numAtoms$; $i$++) **do**
 2:     nbrList = getNeighbors($i$);
 3:     **for** (int $j = 0$; $j < len$(nbrList); $j$++) **do**
 4:         $k$ = nbrList[$j$];
 5:         gEnergy += computeE(atoms[$i$], atoms[$k$]);
 6:         $f_{i,k}$ = computeF(atoms[$i$], atoms[$k$]);
 7:         gForce[$i$] += $f_{i,k}$;
 8:         gForce[$k$] -= $f_{i,k}$;
 9:     **end for**
10: **end for**
---

As outer loops of interaction functions (line 1 in Alg. 1) were identified to be the targets for multi-threading, these loops were made OpenMP parallel by dividing the atoms among threads, and certain local variables were made thread private. On an architecture with shared memory parallelism, this situation creates race conditions on the global force data structure, as atoms assigned to different threads may be neighbors of each other or may have common neighbors. To eliminate race conditions, while ensuring a balanced workload distribution among threads, we experimented with different thread parallelization strategies which are portable to compilers with OpenMP support. It should be noted that advanced thread parallel algorithms based on spatial partitioning among threads have recently been proposed for achieving good performance on multi/many-core architectures [27, 28, 29]. Due to the wide variety of kernels involved in ReaxFF and the non-trivial challenges associated with resolving the race conditions in dynamic bond, angle and torsion interactions and QEq solvers, our current efforts have focused on relatively simple strategies that we summarize below. Further optimizations in the spirit of the more advanced approaches are planned as future work.

**Critical Regions:** In this implementation, each thread computes the energy and forces corresponding to an interaction assigned to it. The updates to the `gEnergy` and `gForce` data structures of Alg. 1, which are shared by all threads, were enclosed within OpenMP critical directives to avoid race conditions. Incurring thread locks via the critical regions within the inner loops was observed to be very inefficient due to the increasing overhead of using the lock. This eroded most of the performance gains in our test systems when using more than a couple of threads.

**Transactional Memory:** As an alternative thread parallelization strategy, we experimented with transactional memory. Although not a part of the OpenMP standard, hardware implementations compatible with OpenMP are available as transactional memory extensions on recent Intel chips supporting the IA64 architecture and as transactional memory (TM) atomics for IBM systems via the XLC compiler.

For this work, we explored the usage of TM atomics (tm_atomic directive) nested within OpenMP parallel regions on Blue Gene/Q. A typical TM implementation consists of essentially replacing OpenMP critical directives with `tm_atomic` in the application code, and passing `-qtm` on the command line during compilation. Blue Gene/Q implements TM support at the hardware level within the L2-cache by tracking memory conflicts for the atomic transaction group. If conflicts are found, an atomic transaction-level rollback is executed, which restores the state of the memory, and the atomic transaction is retried a limited number of times before a lock is imposed and the code is serialized through the atomic region. In this fashion, multiple threads can execute the code on shared memory data concurrently without incurring the overhead of locks. However, there are performance factors to be considered. There is a certain performance overhead in generating the atomic transaction each time it executes, and significant overhead can be incurred if a conflict is found and a rollback occurs. So, the key to TM performance is to have a significant amount of work in the transaction while avoiding frequent conflicts with other threads. There are runtime environment variables supported by the XLC compiler that tell the application to generate reports detailing the runtime characteristics of the transactions. These reports can give clues regarding the impact of TM on performance and be used to guide further tuning of the code.

This approach was performed in several of the energy and force computation functions in ReaxC-OMP, where a tunable number of iterations in the inner-loop pairwise computations was chunked together into one transaction. However, no significant performance improvement over the baseline OpenMP implementation, where race conditions were resolved using *critical* sections, could be attained with any number of iterations. With a small number of iterations in a chunk, there were few conflicts but a lot of transactions, so the transaction generation overhead prevented any speedup. When the chunk size was increased, there were larger but fewer transactions. In this case, the increased number of conflicts resulted in a significant number of rollbacks which again prevented any speedup. In these computations, the TM conflicts arose because disparate threads were executing pairwise computations with common neighbors based on the division of labor occurring on the outer atomic index loop, but the atomic index has little correlation with spatial decomposition.

**Data Privatization:** To prevent race conditions, we also explored the use of thread-private arrays for force updates and OpenMP reductions for energy updates. We first discuss the rationale for this choice and note specific implementation issues regarding each interaction later in this section and the next one.

In this scheme, instead of a thread updating the `gForce` data structure directly at the inner loop level, each thread is allocated a private force array at the start of the simulation which it updates independently during force computations. After all force computations are completed, thread-private force arrays are aggregated (reduced) into the `gForce` array to compute the final total force on each atom. Despite the performance overhead of this additional reduction step, the data-privatization methodology was much more efficient than thread locks and scaled well with large numbers of threads (up to 16 as demonstrated in the performance evaluation section).

Finally, in our OpenMP implementation, system energy tallies are handled with relatively little performance overhead via the OpenMP reduction clause at the outer loop level. Additionally, electrostatic and virial forces need to be tallied for each pairwise interaction. The original MPI-only implementation utilized pre-existing serial functions within the pair-wise force field base class (Pair) in LAMMPS for this purpose. Now, the threaded versions are utilized within the LAMMPS/USER-OMP package according to a methodology consistent with other threaded force field implementations, which substitute the serial setup, tally, and reduction functions appropriately in place.

## 3.2 Thread Scheduling

In OpenMP, static scheduling is the default work partitioning strategy among threads. In our outer loop parallelization scheme described above, static scheduling would partition $n$ atoms into $t$ chunks ($t$ being the number of threads) consisting of approximately $\frac{n}{t}$ contiguous atoms in the list. While static scheduling incurs minimum runtime overheads, such a partitioning may actually lead to load imbalances because some atoms may have a significantly large number of interactions in comparison to others in a system where atoms are not distributed homogeneously throughout the simulation domain. Also, some atoms may be involved in a large number of 3-body, 4-body and hydrogen bond interactions, while others may have none due to specifics of the ReaxFF model and relevant chemistry. As an illustrative example, a plot of the assignment of candidate valence angle interactions to atoms from the LAMMPS/FeOH3 example on a single process is shown in Figure 4. In LAMMPS, the atom list is reordered based on spatial proximity to improve cache performance, and in this particular case, the majority of valence angle interactions involve atoms appearing at the beginning of the atoms list. The remaining atoms (which corresponds to more than 80% of atoms in the system) do not own any angle interactions. As a typical simulation progresses and atoms move, the ordering within the atom list may change, but the imbalance of per-atom work would remain. Therefore statically scheduling the work across threads in contiguous chunks of the atom list can degrade performance as some threads can own considerably more angles than the average number. This example focused on valence angle interactions, but similar workload distributions exist for other interactions due to the chemical nature of the species simulated, making this a general issue that needs to be addressed.

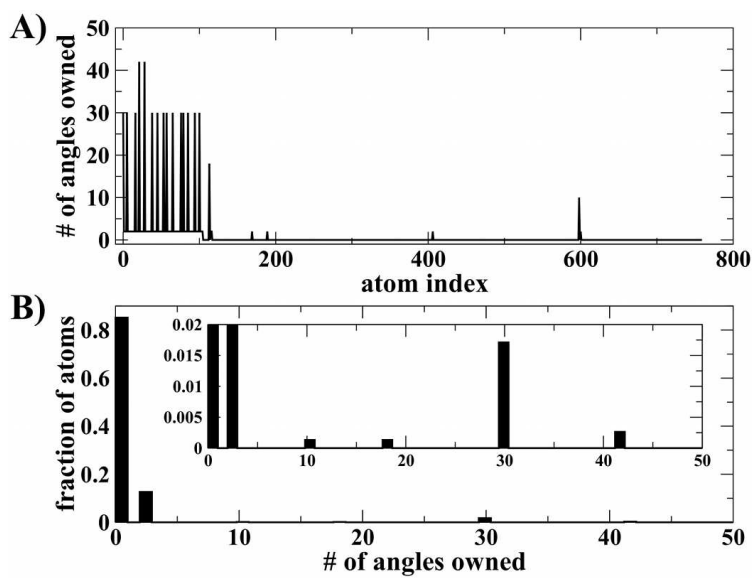

Figure 4: A) How the assignment of valence angle interactions to atoms generates an imbalanced distribution for LAMMPS FeOH3. B) Fraction of atoms with specified count of angles owned. The inset shows a magnified view of the fraction of atoms with the majority of assigned work. This is one example where a naive assignment of work to threads is inefficient and degrades performance when scaling to a large number of threads.

For the majority of cases, the use of the dynamic scheduling option in OpenMP was found to ensure a good balance of work among threads as opposed to explicitly assigning per-thread work beforehand. However, there exists an important trade-off regarding the chunk size. Smaller chunks are better for load balancing, but they may incur significant runtime overheads (default chunk size for dynamic scheduling is 1). Larger chunks reduce scheduling overheads, but with larger chunks load balancing is harder to achieve and the number of tasks that can be executed concurrently decreases. In the new ReaxC-OMP package, we empirically determined the scheduling granularity. For example, comparing the performance for the 16.6 million particle benchmark on 8,192 BG/Q nodes, a chunksize of 20 atoms gives slightly better performance using 8 MPI ranks and 8 OpenMP threads per rank on each node (Figure 5). As the number of threads per MPI rank increases (and number of MPI ranks per node decreases), a chunksize of 25 was found to be optimal. For the chunksizes sampled in the range 10-50, a maximum deviation of 6% in performance was observed relative to the performance with the smallest chunksize when running 64 threads per MPI rank. While this parameter needs to be tuned for ideal performance depending on the simulated system and the architecture, its default value is set to 20. Although the measured performance deviation was not significant enough to justify further tuning work, the trend towards highly parallel many-core systems, e.g. Xeon Phi, provides motivation for future work in this area.

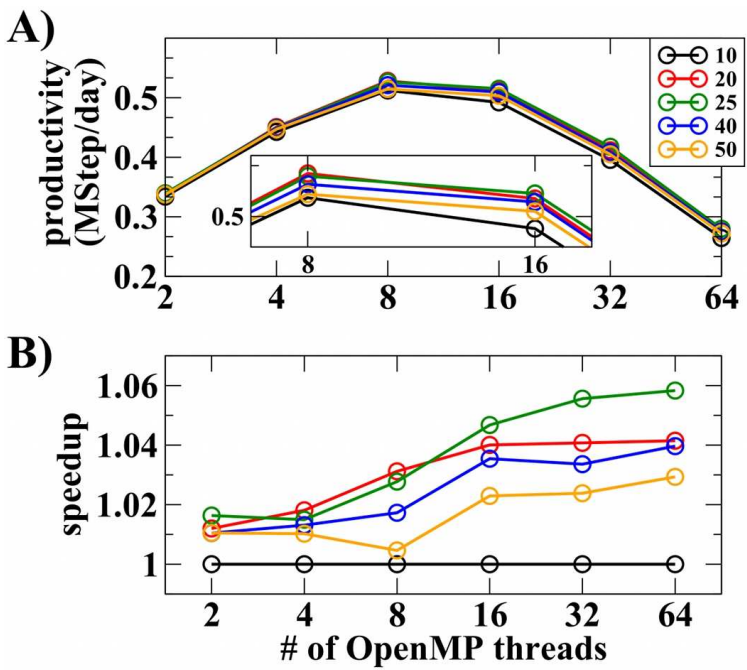

Figure 5: A) Productivity (total simulation steps per time unit) as a function of number of OpenMP threads for a range of chunksizes observed for the 16.6 million particle PETN benchmark on 8,192 BG/Q nodes. B) The observed speedup relative to a chunksize of 10. The inset in A) shows a magnified view for the performance of 8 and 16 threads.

### 3.3 Implementation Details

In this section, we present implementation details regarding key kernels and data structures in ReaxC-OMP.

**Neighbor and Interaction Lists:** Neighbor lists generated by LAMMPS at the request of a force field contain only the neighboring pair information. ReaxC-OMP maintains a separate neighbor list with more detailed information like pair distance and distance vector, as these quantities are needed multiple times during the construction of the bond list and the hydrogen bond list, as well as during force computations. The neighbor list is stored by default as a half list, *i.e.*, for neighboring atoms $i$ and $j$, only a single record is kept. A compact adjacency list format (similar to the compressed row format in sparse matrices) is used for storing the neighbor list.

While a half list is advantageous to reduce the computational and storage costs of the neighbor list, it brings challenges in generating the bond and hydrogen bond lists. Efficient on-the-fly construction of 3-body and 4-body interaction lists requires the bond list to be a full list with both $i$-$j$ and $j$-$i$ bonds available. The hydrogen bond list is generated based on the surrounding atom information of a covalently bonded $H$ atom; but this information needs to be spread throughout a neighbor list that is stored as a half list.

In ReaxC-OMP, we generate the bond and hydrogen bond lists by making a single pass over the neighbor list and, if needed, updating the bond or hydrogen bond lists of atoms $i$ and $j$ concurrently. As with the force computation kernels, the outer loop sweeping over the neighbor list is thread parallelized. The challenge then is in the inner loop, where race conditions may arise due to updates to the bond or hydrogen bond lists of common neighbors. In both cases, race conditions are prevented by introducing *critical regions* that can be executed by a single thread at any given time. For a thread which needs to update the bond or hydrogen bond list of atom $j$ while processing the neighbors of atom $i$, the critical region only includes the reservation of a slot in the relevant list. Once a slot is reserved, all subsequent bond or hydrogen bond related computations are performed outside the *critical region*. In this way, performance penalties associated with *critical regions* are reduced by limiting them to be very short code sequences. We have found the combination of a half-list to store neighbors and the use of critical regions to give good overall performance on moderate number of threads (up to 16) as discussed in the performance evaluation section.

**Pairwise Interactions:** Bond order correction, bond energy and non-bonded interaction computations (*i.e.* van der Waals and Coulomb interactions) constitute the pair-wise interactions in ReaxFF. As described above, these interactions are made OpenMP parallel at the outer loop level and race conditions are resolved through the use of thread-private force arrays and OpenMP reductions for energies. In Alg. 2, we give a simple pseudo-code description of the van der Waals and Coulomb interactions in the non-bonded force computations to illustrate this idea.

---

**Algorithm 2** Threaded non-bonded pairwise force computation

**Input:** Atom list and positions: atoms
**Output:** Potential energy, forces (partial): gEnergy, gForces

 1: #pragma omp parallel reduction (+:PotEng) {
 2: $tid \leftarrow$ omp_get_thread_num();
 3: PairReaxC–>evThreadSetup($tid$);
 4: #pragma omp for schedule(dynamic)
 5: **for** (int $i \leftarrow 0$; $i < numAtoms$; $i$++) **do**
 6:     nbrList $\leftarrow$ getNeighbors($i$);
 7:     **for** (int $j \leftarrow 0$; $j < len$(nbrList); $j$++) **do**
 8:         $k \leftarrow$ nbrList[$j$];
 9:         $evdW, fvdW \leftarrow$ vdWaals(atom[$i$], atom[$k$]);
10:         $eClmb, fClmb \leftarrow$ Coulomb(atom[$i$], atom[$k$]);
11:         PotEng += ($evdW + eClmb$);
12:         tprivForce[$tid$][$i$] += ($fvdW+fClmb$);
13:         tprivForce[$tid$][$k$] -= ($fvdW + fClmb$);
14:         PairReaxC–>evThreadTally($tid$);
15:     **end for**
16: **end for**
17: PairReaxC–>evThreadReduction($tid$);
18: Reduce tprivForces into gForce array
19: }

---

**Three-body Interactions:** One particular challenge in ReaxFF is the dynamic nature of the three-body interactions list. Whether an atom contributes to a three-body valence angle interaction depends on the molecular identity and the surrounding environment of the atom. As such, not all atoms in a system may be involved in a three-body interaction. Additionally, depending on the nature of the molecular species being simulated, only a subset of atoms in the system are designated as the *central atom* of an angle (e.g., see Figure 4).

The three-body interactions are dynamically formed based on the bonds of *central atoms*; they need to be stored in a separate list because four-body interactions are generated based on the three-body interactions present at a given time step. Storing three-body interaction information is expensive in terms of memory, and the number of interactions per atom can vary significantly as shown in Figure 4. Therefore, we first identify which angles are present at a time step without storing them. After all angles have been identified, a per-atom prefix sum is computed. The 3-body interactions are then computed and stored using the global array offsets to eliminate memory clashing between threads.

**QEq:** The dynamic bonding in ReaxFF requires the re-distribution of partial charges at every step. LAMMPS/ReaxC uses the the charge equilibration method (QEq) [11, 21] which models the charge re-distribution as an energy minimization problem. Using the method of Lagrange multipliers to solve the minimization problem, two linear systems of equations are obtained with a common kernel $H$, an $N \times N$ sparse matrix where $N$ is the number of atoms. $H$ denotes the coefficient matrix generated from a truncated electrostatics interaction and well-known Krylov subspace methods (CG [30] and GMRES [31]) can be used to solve the charge re-distribution problems [11].

An effective extrapolation scheme that we developed for obtaining good initial guesses and a diagonally preconditioned parallel CG solver yield a satisfactory convergence rate for charge equilibration. This QEq solver had previously been implemented as the `fix qeq/reax` command in LAMMPS. As part of this work, OpenMP threading was applied to several computational loops within the QEq solver, most significantly the sparse matrix vector multiplication (for which the implementation is described in section titled Detailed Performance Analysis) and the construction of the Hamiltonian matrix from the neighbor list. Taking advantage of the fact that the QEq Hamiltonian is symmetric, only unique, non-zero elements of the sparse matrix are computed and stored. Using an atom-based prefix sum, the effort to compute the Hamiltonian matrix is efficiently distributed across threads avoiding potential race conditions to improve performance.

Finally, in the new ReaxC-OMP package, we adopted a concurrent iteration scheme [12] in the Krylov solver that combines the sparse matrix multiplication and orthogonalization computations for the two linear systems involved in charge equilibration. This concurrent iteration scheme helps reduce communication and synchronization (both MPI and OpenMP) overheads.

## 3.4  A Tool for Molecular Species Analysis

As the gap between the processing power and I/O capabilities of HPC systems widens, the conventional way of generating a trajectory output during the simulation and doing a post-processing analysis on this data becomes a major bottleneck. This approach is extremely I/O intensive and does not scale to large atom and/or processor counts. A distinct need in reactive molecular simulations is the analysis of molecular species, which puts even more pressure on the I/O system for a number of reasons. First, each snapshot is rather large, as dynamic bonding information typically requires 100-1000 bytes per atom. Second, to track individual chemical reactions the trajectory output frequency must be higher than the fastest reactive process in the system, even if this process only involves a small subset of all atoms. Third, sub-sampling and time-averaging of bonding information is required in order to distinguish persistent bonds from transient encounters due to thermal and ballistic collisions. Consequently, even for modest system sizes (*e.g.* less than 100,000 atoms and 100 cores), time spent performing I/O and post-processing chemical species analysis can greatly exceed the time spent running the MD simulation itself, thereby significantly hampering the overall productivity. This situation gets exacerbated even further with the performance improvements that we achieve in ReaxC-OMP.

To cope with this problem, we developed a real-time *in situ* molecular species analysis capability integrated within LAMMPS as the `fix reax/c/species` command. This command uses the same spatial decomposition parallelism as the ReaxFF simulation, taking advantage of the distributed data layout, and achieves comparable scaling performance to the MD simulation itself. Bonds between atoms, molecules and chemical species are determined as the simulation runs, and concise summary that contains information on the types, numbers and locations of chemical species is written to a file at specific time steps. As a result, users are now able to monitor the chemical species and chemical reactions in *real time* during large-scale MD simulations with reactive potentials, instead of analyzing huge trajectory files after the simulations have finished.

The *in situ* molecular species analysis algorithm can be summarized with the following steps:

- A pair of atoms, both with unique global IDs, are deemed to be bonded if the bond order value between the two atoms is larger than a threshold specified for this interaction type (default value is 0.3). A molecule ID, that is the smaller value of the two global IDs, is assigned to the pair of atoms. This process is repeated until every atom has been assigned a molecule ID.

- Sorting is performed for all molecule IDs and molecule IDs are reassigned from 1 to M, where M corresponds to the maximum number of molecule IDs. This number M also indicates the number of molecules in the system.

- Unique molecular species are determined by iterating through each of the molecules and counting the number of atoms of each element. Molecules with the same number of atoms per element are identified as the same species. One drawback of this algorithm is that it does not distinguish isomers.

- Finally, each distinct species and their counts are printed out in a concise summary.

In-situ analysis of physical observables such as bond order, bond lengths and molecular species can be beneficial so long as the analysis time remains a small percentage of the simulation time. In this case, one benefits from the distribution of data structures within the simulation code to compute observables in parallel while data is still in memory. As we show in the performance evaluation section, the analysis through the `fix reax/c/species` command exhibits a similar scaling behavior as the simulation itself and incurs only minimal performance overheads which is a significant advantage over the I/O intensive post-processing method.

Additionally, the physical observables can be stored and averaged to determine bonds between atoms based on time-averaged bond order and/or bond lengths, instead of bonding information of specific, instantaneously sampled time steps. Such a capability can be achived by using the new `fix reax/c/species` command in conjunction with the time-averaging of per-atom vectors function (`fix ave/time`) in LAMMPS.

## 3.5 Verification and Validation through Science Cases

The ReaxC-OMP package has been developed in close collaboration with domain scientists at Sandia and Argonne National Laboratories. The two science cases used for verification and validation included the study of the effects of material defects and heterogeneities in energetic materials, and investigation of graphene superlubricity. For both cases, energies and forces computed at each step, as well as the overall progression of molecular trajectories have been validated against the original MPI-only implementation in LAMMPS. Below we give a brief summary of both studies.

**Energetic Materials:** Material defects and heterogeneities in energetic materials play key roles in the onset of shock-induced chemical reactions and the ignition of hotspots by lowering initiation thresholds. A hot spot with increased temperature/stress and enhanced chemical reactivity was previously observed in a micron-scale, 9-million-atom PETN single crystal containing a 20 nm cylindrical void [32]. Using the new ReaxC-OMP code described in this paper, simulations to model hot spots in PETN crystals were extended from the previous 50 ps mark of Shan and Thompson [32] to the 500 ps mark, which is sufficiently long to estimate the hot spot growth rate and elucidate its mechanism. This study used 8,192 IBM BlueGene/Q nodes on Mira at the Argonne National Laboratory for a total of approximately 400 hours (approximately 50 million core-hours in total). A manuscript discussing the detailed results and findings from this work is under preparation.

**Graphene Superlubricity:** In a separate study, ReaxFF simulations were used to assist with obtaining atomic-level insight into the mechanism for macro-scale superlubricity enabled by graphene nano-scroll formation [33]. Our ReaxC-OMP software facilitated the exploration of large-scale systems under conditions of ambient humidity. These simulations helped shed light on and attribute superlubricity to a significant reduction in the interfacial contact area due to the scrolling of nanoscale graphene patches and incommensurability between graphene scroll and diamond-like carbon [33].

## 4 Performance Evaluation

Performance benchmarks were executed on Mira, a 48-rack IBM Blue Gene/Q system at Argonne. Each compute node on Mira contains a PowerPC A2 processor running at 1.6 GHz with 16 cores and 16 GB RAM. Each core on the A2 processor has 4 hardware threads, yielding a total of 64 threads per node. Mira's 49,152 compute nodes are connected with each other using a proprietary 5D torus interconnection network to provide a peak computing speed of 10 petaflops. Calculations in this study utilized up to 16 BG/Q racks with 1024 compute nodes per rack. All code was compiled using the IBM XL C/C++ compiler (Aug. 2015 version) and linked with the MPICH-based MPI v2.2 (with error checking or asserts turned off) over PAMI (Parallel Active Message Interface) on BG/Q. PAMI uses low-latency and high-bandwidth messaging via shared memory within a node via atomic primitives for lockless queues in L2 (p2p) and shared addressing (collectives). The following optimization flags were used during compilation: `-g -O3 -qarch=qp -qtune=qp -qsimd=auto -qhot=level=2 -qprefetch -qunroll=yes`, and the `-qsmp=omp:noauto` flag was added for compiling the OpenMP enabled ReaxC-OMP code.

The PETN crystal benchmark available on the LAMMPS website was used in performance evaluation studies. Replicas of the PETN system containing up to 16.6 million particles were examined. In all benchmark tests, charge equilibration (QEq) was invoked at every step with a convergence threshold of $10^{-6}$.

**Active Idling:** Since the general threading scheme in ReaxC-OMP consists of several parallel regions independently implemented across disparate functions, the execution path of the code oscillates between threaded and non-threaded regions. The shared memory parallel (SMP) runtime treatment of idle threads could have a significant impact on performance. It is optimal in this case for the threads to continue spinning and remain as active as possible in between the thread-parallel regions, so that when they again have work to do, they can resume efficiently. In OpenMP, this is achieved by setting the runtime environment variable OMP_WAIT_POLICY to be ACTIVE, which is applicable on all platforms with OpenMP support. In addition, BG/Q systems provide the BG_SMP_FAST_WAKEUP environment variable to reduce the thread wake up time which has been set to YES for performance evaluations reported in this paper.

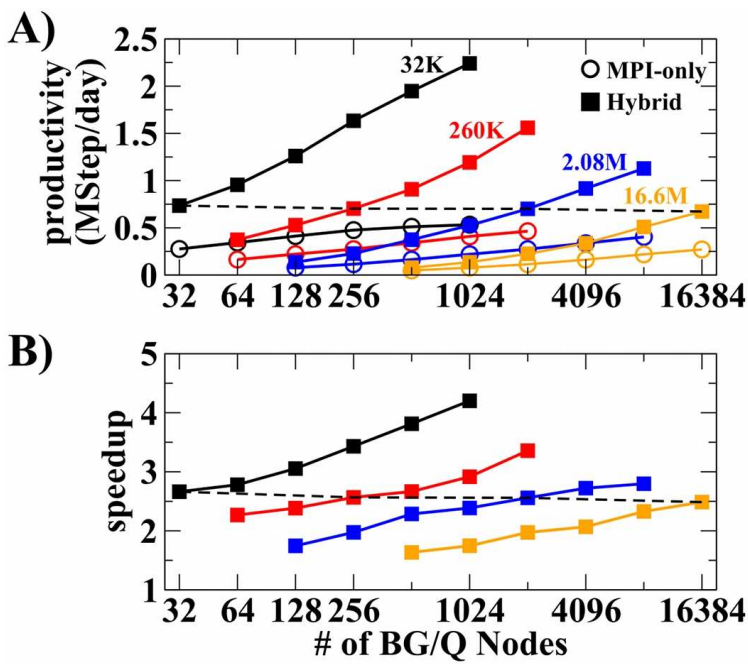

Figure 6: A) Measured productivities of original MPI-only (open circles) and the new hybrid parallel ReaxC-OMP (solid squares) implementations reported as millions of timesteps (MSteps) per day for four different sizes of the PETN crystal benchmark: 32,480 (black), 259,840 (red), 2,078,720 (blue), and 16,629,760 (orange) particles. B) Relative speedups of hybrid vs. MPI-only implementations. The dashed line in each plot is to guide the eye for the weak-scaling comparisons.

### 4.1 Performance and Scalability

To examine the performance as a function of the number of MPI ranks used per node and OpenMP threads used per MPI rank, we performed a benchmark test on the PETN crystal replicated up to 16.6 million atoms. A range of 32 to 16,384 BG/Q nodes (each with 16 cores) was used to best reflect a representative range of HPC resources typically available to a user. Runs using between 1 to 64 hardware threads per node were sampled. Specifically, the number of MPI ranks per node was varied in powers of 2 from 1 to 64 both for the MPI-only and hybrid parallel implementations. Additionally, the number of threads used per MPI-rank were varied in the hybrid code. 100-2000 MD step simulations were performed depending on system size with a standard setup, i.e., 0.1 fs time step size and re-neighboring checked every 10 MD steps. Trajectory files are not written (per the original benchmark available on the LAMMPS website [14]), thus, the timings reported in this section are representative of only the computation and communication costs of the ReaxFF simulations. With support in LAMMPS for MPI-IO and the writing of trajectory files per subset of MPI ranks, the performance costs associated with I/O are expected to be 1-5% of the runtime for these system sizes.

**Performance Improvements:** To quantify the performance improvements achievable with the hybrid parallel ReaxC-OMP package, the original MPI-only ReaxFF implementation in LAMMPS, i.e. the USER-ReaxC package, was used as the baseline case. Performance results for both codes on the PETN crystal benchmark are plotted in Figure 6 with systems ranging from 32 thousand to 16.6 million particles. Each data point in this figure reports the best runtime for all possible MPI-only and MPI/OpenMP configurations for the two codes on each node count. Ultimately, we observed that the best-performing runs use 2-4 hardware threads per core. Performance of the MPI-only runs usually max out at 32 ranks per node, provided that enough memory is available (note that there is only 16 GB of memory per node). For the hybrid parallel runs, optimal performance was typically observed using 8 MPI ranks with 8 OpenMP threads each. In some cases, a configuration of 4 MPI ranks with 16 threads has given slightly better performance (see below for a more detailed discussion). We note that these observations are in-line with expectations from the BG/Q hardware. Clearly, an application needs to execute at least 16 software threads per node, otherwise some cores will be idle. Each core has four hardware threads and a hardware thread can only issue one integer/load/store or floating-point instruction at a time. Therefore to keep the instruction pipelines filled, at least two software threads per core need to run for a total of 32 software threads per node.

| PETN | Simulation Time per Step (s) | Weak Scaling |
|---|---|---|
| 32 K atoms on 32 nodes | 0.118 | 100% |
| 260 K atoms on 256 nodes | 0.123 | 96% |
| 2.08 M atoms on 2,048 nodes | 0.128 | 93% |
| 16.6 M atoms on 16,384 nodes | 0.131 | 91.5% |

Table 1: Weak scaling of the PETN simulation on Mira. The performance of the 32K particle system on 32 nodes was taken as the base case.

Figure 6 shows that for our smallest system (32K atoms), the overall execution time on 1,024 BG/Q nodes for the hybrid code was 4.2 times smaller than that of the MPI-only code. With the larger system sizes (2.08 M and 16.6 M atoms), consistent speedups of $1.5\times$ to $3\times$ (Figure 6b) were observed. Note that the higher speedups achieved with smaller systems is due to the higher communication and redundant computation to useful computation ratio in these systems, as discussed in the motivation for a hybrid implementation section.

**Weak Scaling:** Performance numbers given in Figure 6a show that the hybrid code also exhibits excellent weak scaling efficiency. As shown in Table 1, taking the performance of the 32K particle system on 32 nodes as our base case for the hybrid implementation, we observe a weak scaling efficiency of 96% with 260 K particles on 256 nodes, 93% with 2.08 M particles on 2,048 nodes and 91.5% with on 16.6 M particles on 16,384 nodes. The dashed lines in Figure 6, serving to guide the eye, connect the points used in the weak-scaling analysis. Speedups of about $2.5\times$ over the MPI-only code is observed in each case.

Overall, the productivity (number of steps per day) gains with the original MPI-only code remains modest even on large systems. Conversely, we observed that productivity with the hybrid parallel implementation continues to improve with the usage of more resources (number of nodes). Since one of the main bottlenecks in computational studies using MD simulations is the extremely long wall-clock times needed to reach simulation time-scales where interesting scientific phenomena can be observed (nanoseconds and beyond), from users' perspective, this is a very important capability provided by the hybrid implementation.

|  | **32 Nodes** | | | **1024 Nodes** | | |
|---|---|---|---|---|---|---|
| **Kernel** | **MPI-only** (s) | **Hybrid** (s) | **Speedup** | **MPI-only** (s) | **Hybrid** (s) | **Speedup** |
| Write Lists | 34.7 | 6.5 | 5.3 | 19.6 | 2.3 | 8.5 |
| Init. Forces | 29.8 | 12.1 | 2.5 | 16.7 | 2.6 | 6.4 |
| Bond Orders | 11.3 | 1.8 | 6.3 | 6.3 | 0.53 | 11.9 |
| 3-body Forces | 5.4 | 2.0 | 2.7 | 2.8 | 0.34 | 8.2 |
| 4-body Forces | 3.7 | 2.3 | 1.6 | 1.9 | 0.40 | 4.8 |
| Non-Bonded For | 6.9 | 6.7 | 1.03 | 0.48 | 0.46 | 1.04 |
| Aggregate For | 19.8 | 3.6 | 5.4 | 11.7 | 1.7 | 6.9 |
| QEq | 15.5 | 22.4 | 0.7 | 12.2 | 12.0 | 1.02 |
| Other | 0.27 | 0.69 | 0.4 | 0.07 | 0.17 | 0.4 |
| Total time | 131.3 | 59.3 | 2.2 | 77.9 | 21.8 | 3.6 |

Table 2: Timing breakdown in seconds for the MPI-only and hybrid versions on 32 and 1024 BG/Q nodes for key ReaxFF kernels with the PETN crystal benchmark containing 32 thousand particles. These simulations have been executed for 500 steps. The ideal configuration for the MPI-only calculations used 32 MPI ranks per node, while the ideal configuration for the hybrid version was determined to be 4 MPI ranks per node and 16 OpenMP threads per rank.

## 4.2 Detailed Performance Analysis

Next, we compare the speedups on a kernel basis obtained by the hybrid implementation running with the ideal number of threads over the MPI-only version. In MPI-only simulations, 32 cores out of the 64 available have been used, as we observed that 32 MPI ranks per node yielded better overall performance in comparison to using all available hardware threads. This is likely due to the increased overheads on large number of MPI ranks, as well as limited cache space available on the IBM BG/Q architectures, which has 16 KB private L1 cache per core and 32 MB shared L2 cache. Simulations with the hybrid implementation, however, could fully utilize all the threads using 4 MPI processes with 16 OpenMP threads per node.

Table 2 gives a breakdown of the timings for the key phases in the PETN crystal benchmark containing 32 thousand particles. For this system, the thickness of the ghost region is 10 Å and the size of the simulation box is 66.4 Å x 75.9 Å x 69.9 Å. The kernels that involve significant redundant computations at the ghost regions are *write lists*, which computes neighbor atom information and distances, *init forces*, which initializes the bond and hydrogen bond lists, *bond orders*, *3-body forces*, *aggregate forces* and to some extent *4-body interactions*. Note that most of these kernels are bond related computations. With the hybrid implementation, we observe significant speedups in all these kernels as the hybrid implementation reduces redundancies at process boundaries. On 1024 nodes, the achieved speedups increase even further, as the ratio of ghost region to the actual simulation domain increases considerably when using the MPI-only version.

We do not observe any significant speedup for *nonbonded forces*, which is expected because this kernel avoids redundant computations in ghost regions, as described in the implementation section. Contrary to our expectations though, for the *QEq* kernel, our hybrid implementation has performed worse than the MPI-only execution on 32 nodes (15.5 s vs. 22.4 s), and only slightly better on 1024 nodes (12.2 s vs. 12.0 s). The QEq kernel is an iterative solver consisting of expensive distributed sparse matrix vector multiplications (SpMV) in the form of $Hx_i = x_{i+1}$ followed by a halo exchange of partial charges at each step. The QEq matrix $H$ is a symmetric matrix, and the original MPI-only implementation exploits this symmetry for efficiency. In the hybrid implementation, we opted to continue exploiting the symmetry and resolved race conditions between threads by using private partial result vectors for each thread. Our tests show that this is computationally more efficient than not exploiting the symmetry at all (which would increase SpMV time by a factor of 2), but still does not perform as well as the MPI-only SpMV computations. This is potentially due to increased memory traffic and cache contentions associated with private result arrays (see Figure 8 for details). This performance degradation in SpMV computations takes away the gains from reduced communication overheads achieved with the hybrid implementation. As a result, for smaller node counts, the QEq computations are carried out more efficiently using the MPI-only version.

|  | 64 Nodes | | | 2048 Nodes | | |
| --- | --- | --- | --- | --- | --- | --- |
| **Kernel** | **MPI-only** (s) | **Hybrid** (s) | **Speedup** | **MPI-only** (s) | **Hybrid** (s) | **Speedup** |
| Write Lists | 11.1 | 2.7 | 2.3 | 4.8 | 0.6 | 8.0 |
| Init. Forces | 9.6 | 6.6 | 1.4 | 4.1 | 0.8 | 5.1 |
| Bond Orders | 3.4 | 0.7 | 4.8 | 1.6 | 0.2 | 8.0 |
| 3-body Forces | 1.8 | 0.8 | 2.2 | 0.7 | 0.2 | 3.5 |
| 4-body Forces | 1.4 | 0.8 | 1.8 | 0.4 | 0.1 | 4.0 |
| Non-Bonded For | 5.8 | 4.3 | 1.3 | 0.3 | 0.3 | 1.0 |
| Aggregate For | 5.5 | 1.3 | 4.2 | 2.8 | 0.4 | 7.0 |
| QEq | 7.5 | 7.7 | 0.97 | 2.4 | 2.8 | 0.85 |
| Other | 0.18 | 0.34 | 0.53 | 0.01 | 0.06 | 0.17 |
| Total time | 46.4 | 25.5 | 1.8 | 17.7 | 5.7 | 3.1 |

Table 3: Timing breakdown in seconds for the MPI-only and hybrid versions on 64 and 2048 BG/Q nodes for key ReaxFF kernels with the PETN crystal benchmark containing 260 thousand particles. These simulations have been executed for 100 steps. In its ideal configuration MPI-only calculations used 32 MPI ranks per node, while the ideal configuration for the hybrid version was determined to be 4 MPI ranks per node and 16 OpenMP threads per rank.

Note that the increased memory traffic and cache contention issues are also present in other kernels due to the use of thread-private arrays. However, those kernels perform several floating point operations per force or bond update, and the number of threads in these tests have empirically been optimized for best performance. On the other hand, in SpMV computations, only two floating point operations (multiply and add) are needed for each non-zero matrix element. The relatively low arithmetic intensity of the QEq kernel explains the poor performance obtained in this kernel. Our future work will focus on the development of more efficient SpMV algorithms that eliminate the use of thread-private arrays and are customized based on the sparsity structures of QEq matrices.

In Table 3, we present a similar breakdown for the PETN crystal benchmark with 260 thousand atoms on 64 and 2,048 BG/Q nodes. In this case, the observed speedups on a per kernel basis are relatively lower. Note that, the simulation domain is 8 times larger than that of the 32 thousand atom case, whereas the number of nodes used is only doubled (from 32 nodes to 64 nodes, and 1,024 nodes to 2,048 nodes). Therefore the ratio of the ghost region to the actual simulation domain is lower in this case, resulting in reduced, but still significant, performance gains.

## 4.3 Number of MPI Ranks and Threads per Rank

Based on Figures 2 and 3, one would expect the best productivity to be achieved using a single MPI process per node and 64 OpenMP threads per process. However, in our tests we observed that the productivity initially increases with the number of threads, but starts decreasing after 8 or 16 threads. A detailed examination of performance with respect to the number of OpenMP threads for the PETN benchmark with 2.1 million particles is shown in Figure 7. Using 4 MPI processes with 16 threads or 8 MPI processes with 8 OpenMP threads per node offers the best performance in the range of 512 to 8,192 nodes for this system. As the number of nodes increases, the improved performance going from 2 to 8 threads is a result of the 4× fewer spatial decomposition domains (and MPI processes) and decreased volume of MPI communication to keep all domains synced at each step in the simulation. In general, we observed that the productivity gains from using hybrid parallelism is more pronounced with increasing node counts.

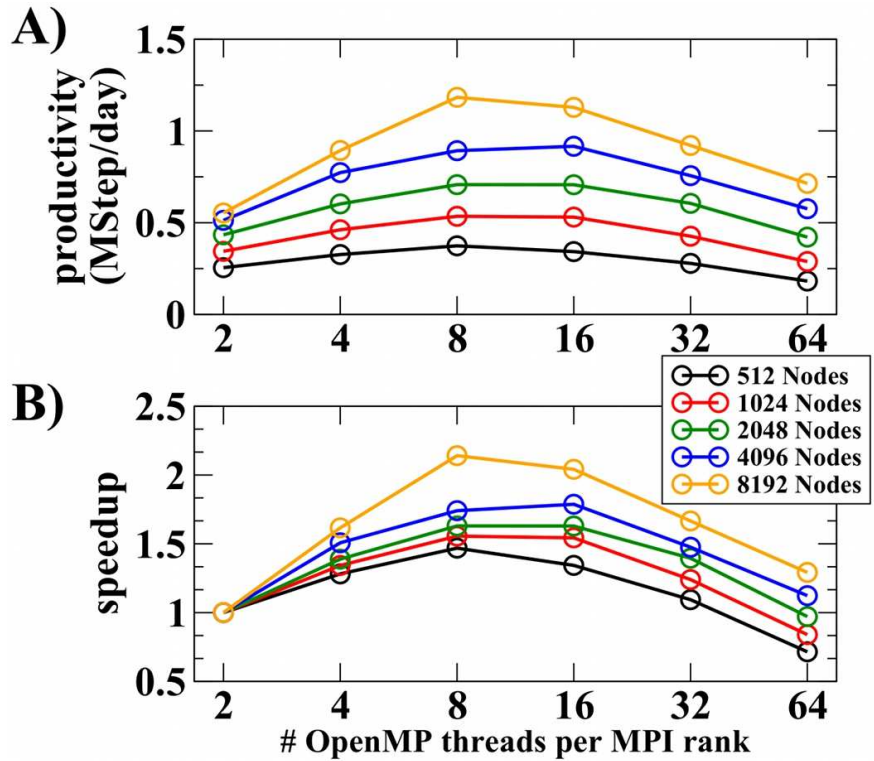

Figure 7: Productivity (A) and parallel speedup (B) of the hybrid ReaxC-OMP package as a function of OpenMP threads per MPI process on 512 to 8,192 BG/Q nodes for a PETN system with 2.1 million particles. In all cases, the number of MPI processes per node times the number of threads per MPI process is 64 which is the total thread count on a BG/Q node.

We believe that the main reason for the limited thread scalability of our approach is the increased memory traffic and cache contention when using a large number of threads. The L1 data and L2 cache hit rates for the QEq SpMV operation were measured for the 32K particle PETN system on 128 BG/Q nodes as shown in Figure 8. For the MPI-only and single-thread hybrid runs, drops in the L1 data cache hit rates are observed when running 2 or 4 software threads per core with a corresponding increase in the L2 cache hit rate. With 16, 32, and 64 OpenMP threads per MPI rank, the L1 data cache hit rate does not exceed 90% and the corresponding L2 hit rate reaches as high as 10-11%, significantly reducing overall performance.

Note that in the hybrid parallel version, we are partitioning atoms to threads using dynamic scheduling for load balancing purposes. This scheme does not necessarily respect data locality, as seen in Figure 8. To take full advantage of the multi-core and many-core parallelism on current and future hardware, our future efforts will focus on improving data locality of workloads across threads. In this regard, the spatial partitioning of the process domain to threads in a load-balanced way, for example by using the nucleation growth algorithm presented by Kunaseth *et al.* [27], is a viable route forward. Also note that unlike classical MD methods where non-bonded computations are the dominating factor for performance, ReaxFF contains a number of computationally expensive kernels such as bond interactions, dynamic 3-body and 4-body lists and hydrogen bonds. To expose a high degree of parallelism and improve thread scalability, we will explore the use of separate teams of threads that asynchronously progress through these key phases of the ReaxFF calculation.

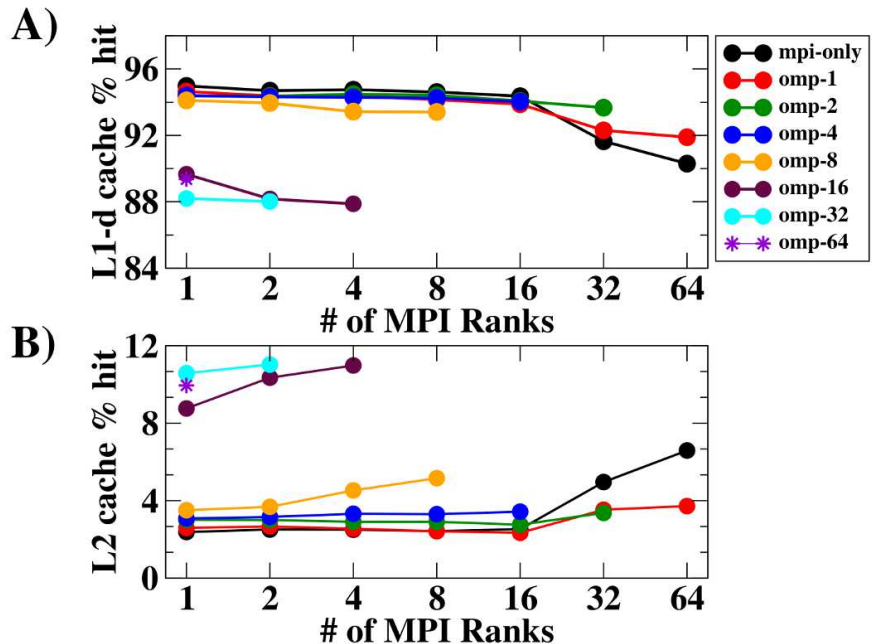

Figure 8: BG/Q L1-d (A) and L2 (B) cache hit rates of QEq SpMV operation for 32K particle system on 128 BG/Q nodes as a function of MPI ranks per node for the MPI-only and hybrid MPI-OpenMP implementations (omp-$N$) for $N$ OpenMP threads.

|                   |          | # OpenMP Threads | | | | | |
| ----------------- | -------- | ----- | ----- | ----- | ----- | ----- | ----- |
| #HW Threads Used  | MPI-only | 2     | 4     | 8     | 16    | 32    | 64    |
| 16                | 1.183    | 1.009 | 0.700 | 0.497 | 0.365 | –     | –     |
| 32                | 2.161    | 1.798 | 1.341 | 0.885 | 0.633 | 0.470 | –     |
| 64                | 4.029    | 3.329 | 2.382 | 1.595 | 1.126 | 0.824 | 0.677 |

Table 4: Per-node heap memory utilization in gigabytes for 32K particle system on 128 BG/Q nodes for the MPI-only and hybrid MPI-OpenMP implementations. The number of software (SW) threads is the product of number of MPI ranks and OpenMP threads.

### 4.4 Memory Overheads due to Data Privatization

One potential drawback to the data privatization approach is the increased memory needs due to duplicating the force array on each thread. For a simple force field, this approach might incur a significant memory overhead overall, if the number of threads is large. In ReaxFF though, the data structures that require major memory space are the neighbor, bonds, 3-body and hydrogen bond lists. In these lists, the number of interactions per atom may range from tens to hundreds, and as we discuss in the implementation details, there is no duplication of these data structures in ReaxC-OMP. In comparison, the force array only stores the force on an atom in $x$, $y$, and $z$ dimensions (i.e., 3 double precision numbers). So under typical simulation scenarios, the duplication of force arrays are not likely to cause significant overheads in terms of the overall memory usage.

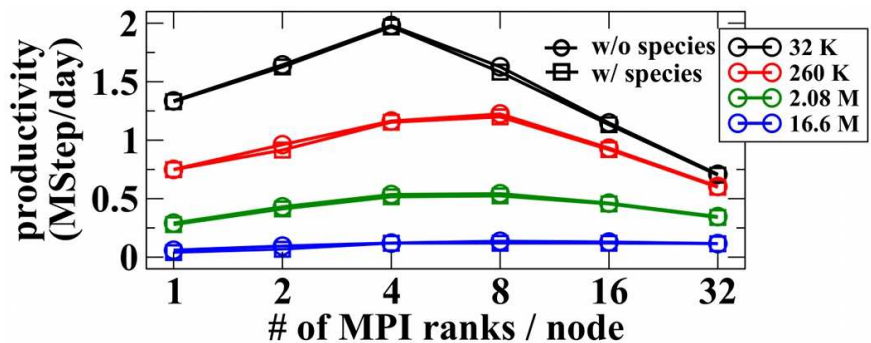

Figure 9: Performance of hybrid MPI/OpenMP implementation both without (circles) and with (squares) the real-time molecular species analysis as a function of MPI processes per node and OpenMP threads per MPI process on 1,024 BG/Q nodes for a PETN crystal containing 32K (black), 260K (red), 2.08 M (green), and 16.6 M (blue) particles. In all cases, the product of MPI ranks per node and OpenMP threads per rank equals 64.

This is illustrated in Table 4 where memory utilization for the MPI-only vs. hybrid versions is reported for a total of 16, 32, and 64 threads. An important observation here is that when utilizing the same amount of computational resources, *i.e.*, the same number of hardware threads per node, the hybrid MPI/OpenMP implementation allocates less per-node memory than the MPI-only version. For example, comparing the MPI-only version with 32 ranks per node and hybrid version with 32 OpenMP threads and a single MPI rank, we observe that the hybrid version requires 4× less memory than the MPI-only version. In its ideal configuration, *i.e.*, using 8 OpenMP threads with 8 MPI ranks or 16 OpenMP threads with 4 MPI ranks, we observe that the hybrid code utilizing all hardware threads allocates roughly the same amount of per-node memory as the MPI-only version running with only 16 MPI ranks (one per core). Despite the duplication of force arrays, the memory space reductions in the hybrid version are the direct consequence of the reduced ghost region ratio, as there are less redundancies in the pairwise, three-body interactions and QEq matrices.

### 4.5 Performance and Scaling with Molecular Species Analysis

Computational expense of the real-time molecular species analysis is illustrated in Figure 9 for all four PETN system sizes on 1,024 BG/Q nodes. In these benchmark tests, bond order values between pairs of atoms are stored every 10 MD steps and subsequently averaged every 1000th step where all molecular species are then written to output in a concise format. On the scale of the plot, the overhead from the molecular species analysis with these settings is on the order of the line thickness amounting up to a maximum of a 5% slowdown for the three smaller systems with overhead increasing with respect to system size. For the largest 16.6 million particle system, the overhead is in the range of 2-35% increasing with the number of OpenMP threads. On a more general note, a 5-25% reduction in productivity is typically observed with the real-time molecular species analysis on other node counts and machines. In comparison, post-processing a large volume of trajectory files with a serial analysis code can be orders of magnitude more costly and time-consuming after accounting for precious simulation time spent writing large trajectory files at high frequency. Overall, the molecular species analysis tool introduced in this study is expected to yield even further productivity gains for the users of the ReaxC-OMP package.

## 5  Related Work

The first implementation of ReaxFF is due to van Duin *et al.* [34]. After the utility of the force field was established in the context of various applications, this serial implementation was integrated into LAMMPS by Thompson *et al.* as the REAX package [35]. Nomura *et al.* have reported on the first parallel implementation of ReaxFF [36] and Nakano *et al.* describes an MPI/OpenMP hybrid version of their code [37], but this codebase remains private to date.

The widely used LAMMPS/User-ReaxC package and the new LAMMPS/ReaxC-OMP package described here are based on the PuReMD code developed by Aktulga *et al.* [13]. The PuReMD codebase contains 3 different packages to ensure architecture portability: sPuReMD [11], PuReMD [12] and PuReMD-GPU [38]. sPuReMD, a serial implementation of ReaxFF, introduced novel algorithms and numerical techniques to achieve high performance, and a dynamic memory management scheme to minimize its memory footprint. Today, sPuReMD is being used as the ReaxFF backend in force field optimization calculations [18] where fast serial computations of small molecular systems are crucial for extending the applicability of the Reax force field to new chemical systems. PuReMD is an MPI-based parallel implementation of ReaxFF, and exhibits excellent scalability. It has been shown to achieve up to $5\times$ speedup over the LAMMPS/Reax on identical machine configurations. PuReMD code has been integrated into LAMMPS as the USER/ReaxC package [14], which actually constitutes the MPI-only version used for comparisons in this study. Acceleration of ReaxFF simulations through the use of GPUs have also been explored recently. Zheng *et al.* report a single GPU implementation of ReaxFF, called GMD-Reax [39]. PuReMD-GPU, a GP-GPU implementation of ReaxFF, achieves a $16\times$ speedup on an Nvidia Tesla C2075 GPU over a single processing core (an Intel Xeon E5606 core).

ReaxC-OMP implementation reported in this study is underway to be released as part of the USER-OMP package in LAMMPS. The Kokkos package, an actively developed C++ library with support for parallelism across different many-core architectures including multi-core CPU, GPGPU, and Intel Xeon Phi, has also recently been released. Our initial investigations show that ReaxC-OMP performs similar or better performance in our limited benchmarking studies, and we plan to do a more detailed performance comparison between the two codes, as part of our future work.

Several *in-situ* analysis and visualization frameworks [40, 41, 42, 43, 44] have been developed for scientific simulations to address the productivity issue with the conventional way of doing post-processing on large datasets. However, analysis of molecular species is a distinct need for reactive molecular simulations. To the best of our knowledge, our integrated tool for molecular species analysis represents the first tool to enable *in-situ* analysis for reactive MD simulations.

## 6  Conclusions

We presented a hybrid MPI-OpenMP implementation of the ReaxFF method in the LAMMPS simulation software and analysis of its performance on large-scale simulations and computing resources. On Mira, a state-of-the-art multicore supercomputer, we observed significant improvements in the computational performance and parallel scalability with respect to the existing MPI-only implementation in LAMMPS. We also presented the implementation and performance results of a tool for in-situ molecular species analysis tailored for reactive simulations. While performance results obtained using a large number of OpenMP threads (e.g. 64) have exhibited limited gains, the threading model employed in this work serves as a useful starting point for extending the thread scalability even further (e.g. to many-core architectures like Intel Xeon Phi). The current hybrid implementation, however, has already proven invaluable in a couple studies involving large-scale, multi-million particle simulations on leadership computing resources. It is expected that a wide community of researchers will have similar successes in their own fields of study as a result of this effort, and the performance benefits will be improved further through future work.

# Acknowledgements

The work by HMA and KAO was partially supported by the National Science Foundation under Grant No. ACI-1566049. TRS acknowledges Oleg Sergeev of VNIIA for fixing several bugs in the LAMMPS' implementation of the real-time molecular species analysis. The authors would also like to thank Dr. Nichols A. Romero of ALCF for useful technical discussions, careful reading of the manuscript and helpful comments. An award of computer time was provided by ALCF's Director's Discretionary program. This research used resources of the Argonne Leadership Computing Facility, which is a DOE Office of Science User Facility supported under Contract DE-AC02-06CH11357. Sandia National Laboratories is a multi-program laboratory managed and operated by Sandia Corporation, a wholly owned subsidiary of Lockheed Martin Corporation, for the U.S. Department of Energy's National Nuclear Security Administration under contract DE-AC04-94AL85000.

# References

[1] MacKerell AD Jr, Bashford D, Bellott M et al. All-atom empirical potential for molecular modeling and dynamics studies of proteins. *J Phys Chem B* 1998; 102(18): 3586–3616.

[2] Cornell WD, Cieplak P, Bayly CI et al. A second generation force field for the simulation of proteins, nucleic acids, and organic molecules. *J Am Chem Soc* 1995; 117(19): 5179–5197.

[3] Jorgensen WL and Tirado-Rives J. The opls potential functions for proteins. energy minimizations for crystals of cyclic peptides and crambin. *J Am Chem Soc* 1988; 110(6): 1657–1666.

[4] Warshel A and Weiss R. An empirical valence bond approach for comparing reactions in solutions and in enzymes. *J Am Chem Soc* 1980; 102(20): 6218–6226.

[5] Knight C and Voth G. The curious case of the hydrated proton. *Acc Chem Res* 2012; 45(1): 101–109.

[6] van Duin ACT, Dasgupta S, Lorant F et al. Reaxff: A reactive force field for hydrocarbons. *J Phys Chem A* 2001; 105: 9396–9409.

[7] Senftle TP, Hong S, Islam MM et al. The ReaxFF reactive force-field: Development, applications and future directions. *Nature PJ Computational Materials* 2016; 2: 15011.

[8] Shan TR, Devine BD, Kemper TW et al. Charge-optimized many-body potential for the hafnium/hafnium oxide system. *Phys Rev B* 2010; 81: 125328.

[9] Liang T, Shan TR, Cheng YT et al. Classical atomistic simulations of surfaces and heterogeneous interfaces with the charge-optimized many body (comb) potentials. *Mat Sci Eng R* 2013; 74: 235–279.

[10] Stuart SJ, Tutein AB and Harrison JA. A reactive potential for hydrocarbons with intermolecular interactions. *J Chem Phys* 2000; 112(14): 6472–6486.

[11] Aktulga HM, Pandit SA, van Duin ACT et al. Reactive molecular dynamics: Numerical methods and algorithmic techniques. *SIAM J Sci Comput* 2012; 34(1): C1–C23.

[12] Aktulga HM, Fogarty JC, Pandit SA et al. Parallel reactive molecular dynamics: Numerical methods and algorithmic techniques. *Parallel Comput* 2012; 38(4-5): 245–259.

[13] Grama A, Aktulga HM and Kylasa SB. PuReMD, Purdue Reactive Molecular Dynamics package. https://www.cs.purdue.edu/puremd, 2014. Accessed on June 8, 2016.

[14] Aktulga HM. LAMMPS/User-ReaxC package, 2010.

[15] Deetz JD and Faller R. Parallel optimization of a reactive force field for polycondensation of alkoxysilanes. *J Phys Chem B* 2014; 118(37): 10966–10978.

[16] Larsson HR, van Duin ACT and Hartke B. Global optimization of parameters in the reactive force field reaxff for sioh. *J Comp Chem* 2013; 34: 2178–2189.

[17] Jaramillo-Botero A, Naserifar S and Goddard WA III. General multiobjective force field optimization framework, with application to reactive force fields for silicon carbide. *J Chem Theory Comput* 2014; 10(4): 1426–1439.

[18] Dittner M, Muller J, Aktulga HM et al. Efficient global optimization of reactive force-field parameters. *J Comput Chem* 2015; : DOI: 10.1002/jcc.23966.

[19] Plimpton S. Fast parallel algorithms for short-range molecular dynamics. *J Comp Phys* 1995; 117: 1–19.

[20] Mortier WJ, Ghosh SK and Shankar S. Electronegativity-equalization method for the calculation of atomic charges in molecules. *Journal of the American Chemical Society* 1986; 108(15): 4315–4320.

[21] Rappe AK and Goddard WA III. Charge equilibration for molecular dynamics simulations. *J Phys Chem* 1999; 95(8): 3358–3363.

[22] Hess B, Kutzner C, van der Spoel D et al. Gromacs 4: Algorithms for highly efficient load-balanced, and scalable molecular simulation. *J Chem Theory Comput* 2008; 4(3): 435–447.

[23] Shaw DE. A fast, scalable method for the parallel evaluation of distance-limited pairwise particle interactions. *Journal of computational chemistry* 2005; 26(13): 1318–1328.

[24] Bowers KJ, Dror RO and Shaw DE. Zonal methods for the parallel execution of range-limited n-body simulations. *J Comp Phys* 2007; 221: 303–329.

[25] Bhatelé A, Kalé LV and Kumar S. Dynamic topology aware load balancing algorithms for molecular dynamics applications. In *Proceedings of the 23rd International Conference on Supercomputing*. ACM, pp. 110–116.

[26] Anderson JA, Lorenz CD and Travesset A. General purpose molecular dynamics simulations fully implemented on graphics processing units. *Journal of Computational Physics* 2008; 227(10): 5342–5359.

[27] Kunaseth M, Richards DF, Glosli JN et al. Analysis of scalable data-privatization threading algorithms for hybrid mpi/openmp parallelization of molecular dynamics. *The Journal of Supercomputing* 2013; 66(1): 406–430.

[28] Pennycook SJ, Hughes CJ, Smelyanskiy M et al. Exploring simd for molecular dynamics, using intel® xeon® processors and intel® xeon phi coprocessors. In *Parallel & Distributed Processing (IPDPS), 2013 IEEE 27th International Symposium on*. IEEE, pp. 1085–1097.

[29] Wu Q, Yang C, Tang T et al. Exploiting hierarchy parallelism for molecular dynamics on a petascale heterogeneous system. *Journal of Parallel and Distributed Computing* 2013; 73(12): 1592–1604.

[30] Hestenes MR and Stiefel E. Methods of conjugate gradients for solving linear systems. *Journal of Research of the National Bureau of Standards* 1952; 49(6).

[31] Saad Y and Schultz MH. GMRES: a generalized minimal residual algorithm for solving nonsymmetric linear systems. *SIAM Journal on Scientific and Statistical Computing* 1986; 7(3): 856–869.

[32] Shan TR and Thompson AP. Micron-scale Reactive Atomistic Simulations of Void Collapse and Hotspot Growth in Pentaerythritol Tetranitrate. *Proc 15th International Detonation Symposium* 2014; in press.

[33] Berman D, Deshmukh S, Sankaranarayanan S et al. Macroscale superlubricity enabled by graphene nanoscroll formation. *Science* 2015; 348(6239): 1118–1122.

[34] Van Duin AC, Dasgupta S, Lorant F et al. Reaxff: a reactive force field for hydrocarbons. *The Journal of Physical Chemistry A* 2001; 105(41): 9396–9409.

[35] Thompson A. LAMMPS/reax package. `http://lammps.sandia.gov/doc/pair_reax.html`, 2009. Accessed on June 8, 2015.

[36] ichi Nomura K, Kalia RK, Nakano A et al. A scalable parallel algorithm for large-scale reactive force-field molecular dynamics simulation. *Comp Phys Comm* 2008; 178(2): 73–87.

[37] Nakano A, Kalia RK, Nomura Ki et al. A divide-and-conquer/cellular-decomposition framework for million-to-billion atom simulations of chemical reactions. *Computational Materials Science* 2007; 38(4): 642–652.

[38] Kylasa SB, Aktulga H and Grama A. Puremd-gpu: A reactive molecular dynamics simulation package for gpus. *Journal of Computational Physics* 2014; 272: 343–359.

[39] Zheng M, Li X and Guo L. Algorithms of gpu-enabled reactive force field (reaxff) molecular dynamics. *Journal of Molecular Graphics and Modelling* 2013; 41: 1–11.

[40] Dorier M, Sisneros R, Peterka T et al. Damaris/Viz: A nonintrusive, adaptable and user-friendly in situ visualization framework. In *LDAV-IEEE Symposium on Large-Scale Data Analysis and Visualization*.

[41] Stone JE, Vandivort KL and Schulten K. Gpu-accelerated molecular visualization on petascale supercomputing platforms. In *Proceedings of the 8th International Workshop on Ultrascale Visualization*. ACM, p. 6.

[42] Dreher M and Raffin B. A flexible framework for asynchronous in situ and in transit analytics for scientific simulations. In *Cluster, Cloud and Grid Computing (CCGrid), 2014 14th IEEE/ACM International Symposium on*. IEEE, pp. 277–286.

[43] Landge AG, Pascucci V, Gyulassy A et al. In-situ feature extraction of large scale combustion simulations using segmented merge trees. In *High Performance Computing, Networking, Storage and Analysis, SC14: International Conference for*. IEEE, pp. 1020–1031.

[44] Hernandez ML, Dreher M, Barrios CJ et al. Asynchronous in situ processing with gromacs: Taking advantage of gpus. In *High Performance Computing*. Springer, pp. 89–106.